\documentclass[useAMS]{mn2e}
\usepackage{times}
\usepackage{graphicx}

\title[{\it XMM-Newton\/} observations of NGC 5204 X-1]{{\it XMM-Newton\/} EPIC observations of the ultraluminous X-ray source NGC 5204 X-1}

\author[T.\,P. Roberts et al.]
{T.\,P. Roberts$^{1,*}$, R.\,S. Warwick$^1$, M.\,J. Ward$^1$,
M.\,R. Goad$^{1,2}$ \& L.\,P. Jenkins$^1$ \\ $^1$ X-ray and
Observational Astronomy Group, Dept. of Physics \& Astronomy,
University of Leicester, University Road, Leicester, LE1 7RH\\ $^2$
Dept. of Physics \& Astronomy, University of Southampton, Highfield,
Southampton, Hants., SO17 1BJ\\ $^*$E-mail: tro@star.le.ac.uk}
\date{}

\def\asca{{\it ASCA~\/}}

\def\xmm{{\it XMM-Newton~\/}}

\def\hst{{\it HST~\/}}
\def\chan{{\it Chandra~\/}}

\def\ergcms{{\rm ~erg~cm^{-2}~s^{-1}}}

\def\ergsec{{\rm ~erg~s^{-1}}}

\def\atpcm{{\rm ~atom~cm^{-2}}}
\def\ctsec{{\rm ~count~s^{-1}}}

\def\H0{{\rm ~km~s^{-1}~Mpc^{-1}}}

\def\kmsec{{\rm ~km~s^{-1}}}

\def\la{\mathrel{\hbox{\rlap{\hbox{\lower4pt\hbox{$\sim$}}}{\raise2pt\hbox{$<$}}}}}
\def\ga{\mathrel{\hbox{\rlap{\hbox{\lower4pt\hbox{$\sim$}}}{\raise2pt\hbox{$>$}}}}}

\def\d25{D$_{25}$}
\def\nh{{$N_{H}$}}

\def\.25{0.25 keV\thinspace}
\def\lx{L$_{\rm X}$}

\begin{document}

\maketitle

\begin{abstract}
We present the results of two \xmm observations of the ultraluminous
X-ray source (ULX) NGC 5204 X-1.  The EPIC spectra are well-fit by the
standard spectral model of a black-hole X-ray binary, comprising a
soft multi-colour disc blackbody component plus a harder power-law
continuum.  The cool ($kT_{in} \sim 0.2$ keV) inner-disc temperature
required by this model favours the presence of an intermediate-mass
black hole (IMBH) in this system, though we highlight a possible
anomaly in the slope of the power-law continuum in such fits.  We
discuss the interpretation of this and other, non-standard spectral
modelling of the data.

\end{abstract}


\section{Introduction}

Ultraluminous X-ray sources (ULXs) are luminous (\lx $> 10^{39}
\ergsec$), extra-nuclear X-ray sources in nearby galaxies.  Their
nature has been the subject of much recent debate (see Miller \&
Colbert 2004 for a review).  In brief, the extreme X-ray luminosities
of many ULXs cannot be explained by stellar-mass black holes radiating
isotropically at (or close to) their Eddington limit, leading to the
suggestion that ULXs harbour a new $10^2 - 10^4$ M$_{\odot}$
``intermediate-mass'' class of black holes (IMBHs; Colbert
\& Mushotzky 1999).  This hypothesis has gained strength from the
recent detection of the soft X-ray emission characteristic of ``cool''
accretion discs, the signature of accretion onto an IMBH, in the X-ray
spectra of several ULXs (e.g. Miller et al. 2003; Miller, Fabian \&
Miller 2004).  However, there is strong circumstantial evidence - in
particular the association of large populations of ULXs with very
active star formation - that suggests that many ULXs may instead be a
type of stellar-mass X-ray binary that somehow exceeds its Eddington
limit (King 2004).  Models for achieving super-Eddington luminosities
from these systems include the ``slim'' disc model (e.g. Ebisawa et
el. 2003), radiation-presure dominated discs (Begelman 2002),
mildly-anisotropic radiation patterns from sources emitting at or
below the Eddington limit (King et al. 2001) and relativistic beaming
(K{\"o}rding, Falcke \& Markoff 2002).

The best testing grounds for our models of ULXs are the nearest and
brightest examples of this class of object.  One such source is NGC
5204 X-1. It has a typical X-ray luminosity of the order $2 - 6 \times
10^{39} \ergsec$ (0.5 -- 8 keV), and is located $\sim 15$ arcsec from
the center of a nearby ($d = 4.8$ Mpc) Magellanic-type galaxy (Roberts
\& Warwick 2000).  It is most notable for possessing a stellar optical
counterpart identified from its accurate \chan position (Roberts et
al. 2001), which \hst observations resolved into two sources plus a
third candidate counterpart (Goad et al. 2002).  Further \hst UV
spectroscopy has now identified HST-3 as the ULX counterpart.  Its
spectral type is similar to a B0 Ib supergiant, but with peculiarities
consistent with some Galactic X-ray binaries (Liu, Bregman \& Seitzer
2004). \chan showed its X-ray spectrum to be a steep power-law
continuum (with $\Gamma \sim 2.8 - 3$), and confirmed that it has been
persistently X-ray luminous for 20 years, with flux variations of no
more than a factor $\sim 3$ (Roberts et al. 2004).  The B0 Ib
counterpart confirms that this ULX is a high-mass X-ray binary, though
there are currently no constraints upon the mass of the putative black
hole in this system.  We present new observations by the \xmm
observatory (Jansen et al. 2001), that provide the most detailed view
of the X-ray spectrum of this ULX, and hence the best constraints upon
the accretor, to date.

\section{The data}

\begin{figure}
\centering
\includegraphics[width=6.4cm]{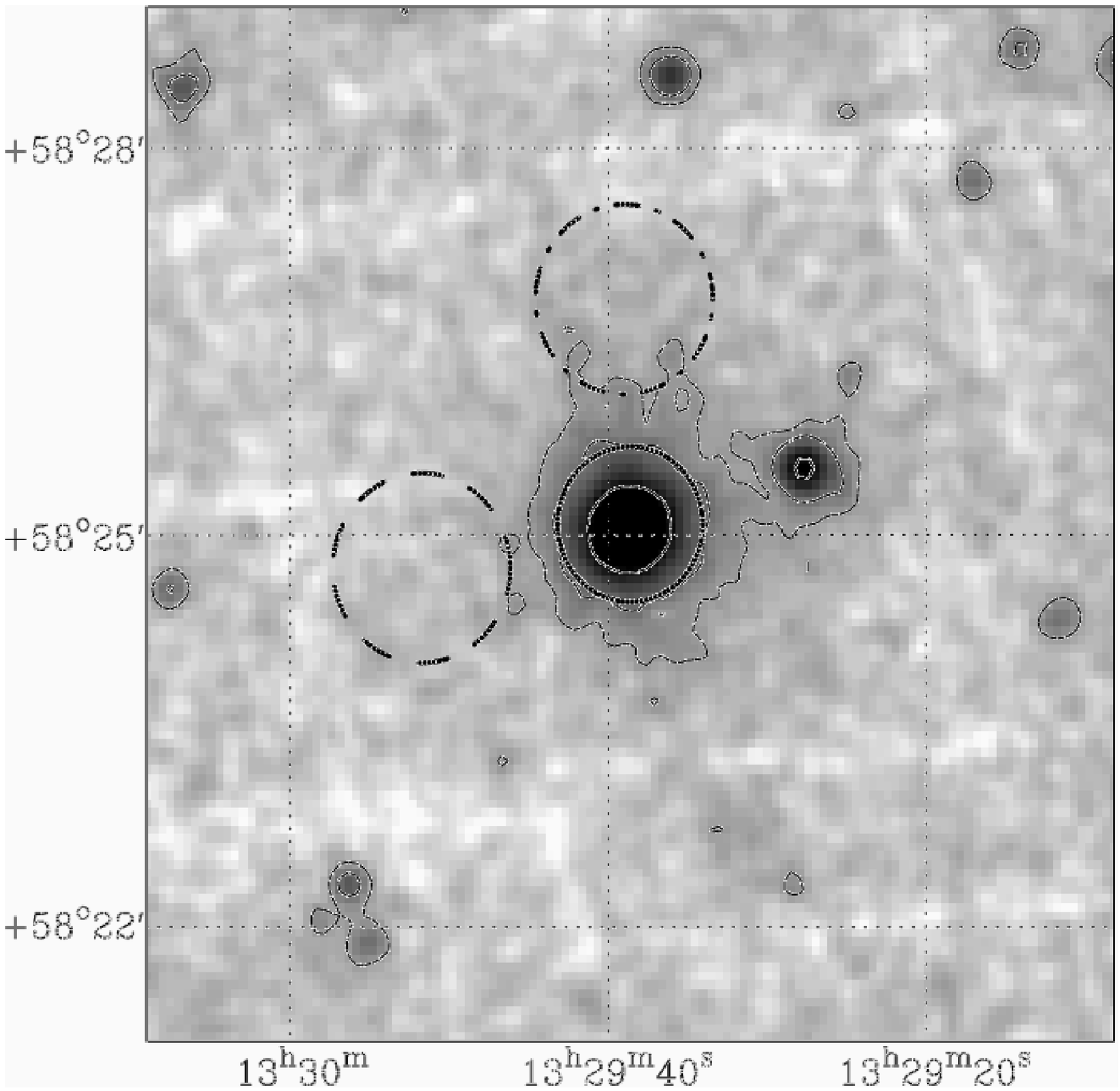}\vspace*{0.4cm}
\includegraphics[width=6.4cm]{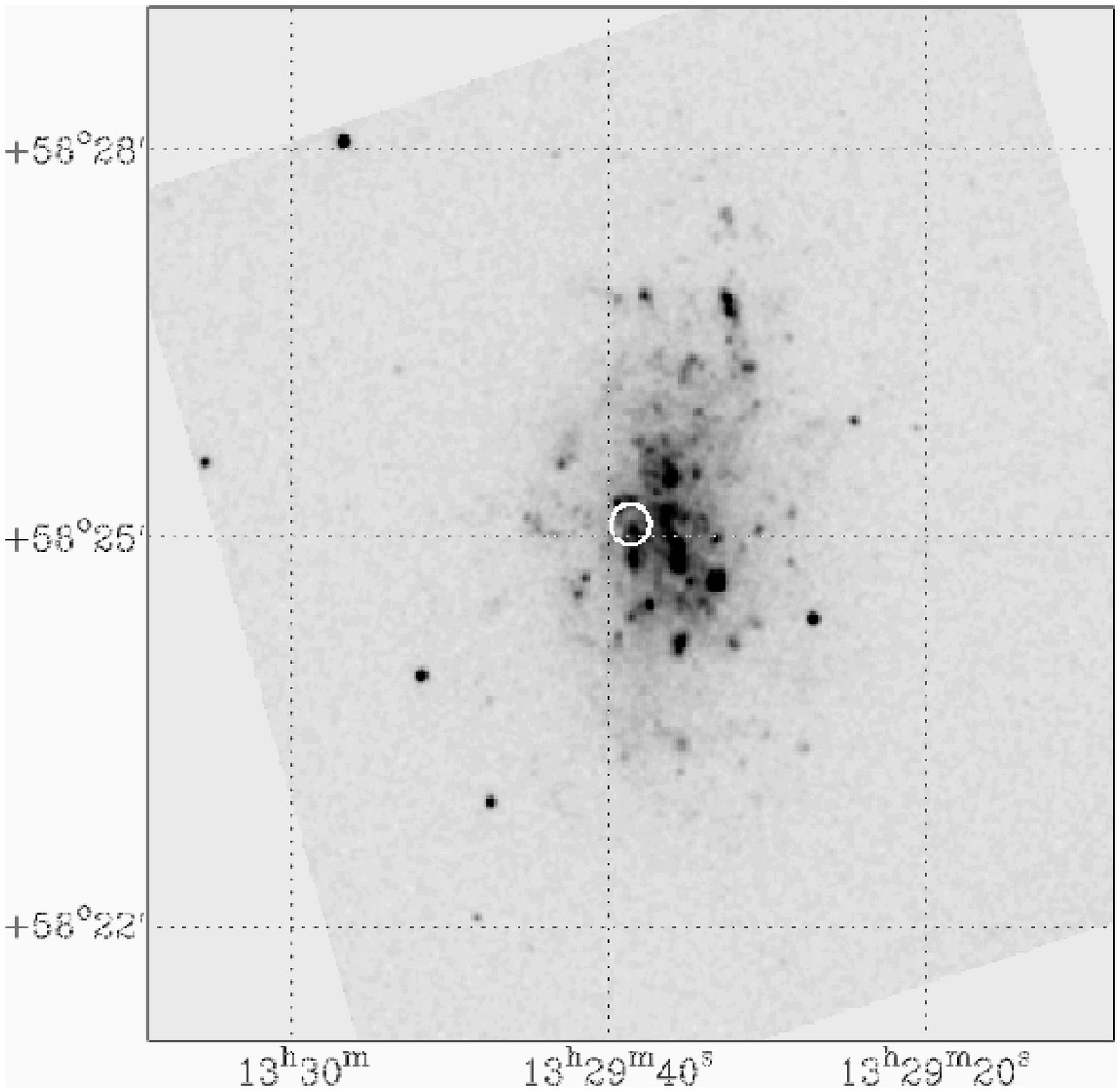}
\caption{({\it top\/}) The central regions of the combined \xmm EPIC
image of NGC 5204.  The image is the sum of the data from all three
detectors and both epochs (subject to the event selections discussed
in the text).  An approximate exposure correction has been
implemented.  The image size is an $8 \times 8$ arcminute$^2$ box, the
data are convolved with a 1 pixel ($\equiv 4$ arcsec) HWHM 2-D
Gaussian mask, and the intensity logarithmically-scaled, for
illustrative purposes.  Contours are overlayed at 5 and 10 count
pixel$^{-1}$.  The data extraction region for NGC 5204 X-1 is shown by
the solid circle; background extractions were made in the dashed
(first observation) and dot-dashed (second observation) circles
respectively.  ({\it bottom\/}) The corresponding OM UVW1 data from the
first observation.  The data have been astrometrically corrected using
reference stars.  The position of NGC 5204 X-1 is highlighted by the
10-arcsec radius circle.}
\label{prettypic}
\end{figure}

The new \xmm data were obtained in two observations separated by a gap
of 109 days.  We present details of the observations in
Table~\ref{obsdet}.  All three EPIC detectors were operated in full
frame imaging mode with the medium filter in place.  Data for the
analysis were extracted from the pipeline product event lists using
the appropriate \xmm {\small SAS v.5.4.1} tools.  The pn data were
filtered to leave only events with flag = 0 and pattern $\leq 4$; MOS
data were filtered for pattern $\leq 12$ and the ``{\tt
$\sharp$xmmea\_em}'' flag.  We initially created images from these
events in the 0.3 -- 10 keV range, and checked the background for
flaring by producing a light curve from the whole detector excluding
the target source.  The background count rate in the first observation
was consistently low ($< 10 \ctsec$ in the pn detector), so no time
filters were applied.  However, the second observation was heavily
contaminated by background flaring, so time intervals during which the
background rate in the pn exceeded 30 $\ctsec$ (in 100 s bins) were
excluded.  This resulted in the loss of $\sim 50\%$ of the data.  The
remaining exposure times (as listed in the GTI file extensions for the
on-axis CCDs, which are also corrected for telemetry drop-outs) are
given in Table~\ref{obsdet}.  The filtered and co-added 0.3 -- 10 keV
images are displayed in Figure~\ref{prettypic}.  Contemporaneous OM
data (in the UVW1, UVM2 and UVW2 filters) were also obtained.  The
UVW1 image of NGC 5204 is displayed in Figure~\ref{prettypic} for
comparison with the EPIC image (although we defer a full analysis of
the OM data to a later work).

\begin{table}
\caption{Details of the \xmm observations.}
\begin{tabular}{lcccccc}\hline
Observation ID & Date & \multicolumn{2}{c}{Useful exposure (s)} \\ &
(yyyy-mm-dd) & pn & MOS \\\hline 0142770101 & 2003-01-06 & 17047 &
18665 \\ 0142770301 & 2003-04-25 & 8609 & 8645 \\\hline
\end{tabular}
\label{obsdet}
\end{table}

NGC 5204 X-1 was detected at its expected position ($13^h29^m38.6^s,
+58^{\circ}25'06''$ (J2000); see e.g. Roberts et al. 2004).  Source
spectra and lightcurves were extracted in a 9-pixel ($\equiv 36$
arcsec) radius circle centred on the source, chosen to sit inside the
chip gaps close to the pn aimpoint.  An 11-pixel radius region centred
$\sim 2$ arcmin from NGC 5204 X-1, on the same pn chip and at
approximately the same distance from the readout nodes, was used to
produce background data (see Figure~\ref{prettypic}).  The spectral
products, including RMF and ARF matrices (which contain the
appropriate corrections for the source extraction region size), were
created using the {\small ESPECGET} tool, and the source spectra were
binned to a minimum of 20 counts per bin before analysis in {\small
XSPEC v11.3}.

\section{Results}

Before embarking upon a detailed analysis of the source properties, it
is worth clarifying whether the source signal is contaminated by other
X-ray emitters within NGC 5204.  In particular, \chan observations of
NGC 5204 X-1 highlight a second point-like source located $\sim 15$
arcsec to the south-east of NGC 5204 X-1 (Roberts et al. 2001; Roberts
et al. 2004), well within the data extraction region used in this
analysis.  If this source maintains its historical flux level with
respect to NGC 5204 X-1, where its \chan count rate was typically $\ll
2\%$ of that of the ULX, it will not significantly contaminate the
data.  We checked whether this is true by extracting the radial
profile of NGC 5204 X-1 from the combined EPIC image.  If the fainter
source had brightened considerably then, since it is located $\sim 15$
arcsec from X-1, it should be marginally resolved in the data and
would cause an excess above the expected \xmm point spread function
(PSF) in one or more radial bins.  However, the off-peak drop-off in
the PSF was smooth, implying that contamination from the fainter
source is not a strong effect.  We also note that the apparently
extended emission outside the source extraction region (c.f.
Figure~\ref{prettypic}) can adequately be explained by the $\sim 15\%$
of the source flux in the wings of the PSF that falls outside the
extraction region\footnote{The apparent north-south structure in
Figure~\ref{prettypic}, mirroring the morphology of the galaxy, may in
part be due to the presence of pn chip gaps.  However, some of the
extension on the periphery of the contour may be due to real faint
X-ray emitters in NGC 5204.}.  Hence there is no evidence for any
strong contamination of the NGC 5204 X-1 signal from other X-ray
sources in NGC 5204.

We tested for temporal variability from NGC 5204 X-1 by extracting
light curves from both sets of data.  The mean 0.3 -- 10 keV count rate
from the combined EPIC detectors rose by $\sim 34\%$ between the two
observations (from 0.86 to 1.15 $\ctsec$), extending the historical
pattern of low-amplitude long-term variability from this ULX
(c.f. Roberts et al. 2004, Figure 7).  Using the spectral models
discussed below gives 0.5 -- 2 keV fluxes of $\sim 0.74$ and $1.05
\times 10^{-12} \ergcms$, which sit within the previously-observed
flux range.  However, neither observation showed strong evidence for
short-term variability.  For example, extracting a power density
spectrum in the full 0.3 -- 10 keV band from the longer first
observation reveals only a slight excess of power above the poisson
noise level at low frequencies ($\la 0.01$ Hz).  Binning the data to
100-s intervals allowed this variability to be detected as excess
variance, albeit at a very low amplitude ($\approx 3\%$ of the total
signal).  The second observation showed no similar variability, with a
95\% upper-limit to its excess variance (on the same timescale) of
$\approx 2\%$ of the total signal.  The following analysis therefore
focuses solely on the time-averaged spectrum extracted from each
observation.

\begin{table}
\caption{Spectral fits.}
\begin{tabular}{lcc}\hline
Model parameter	& 2003 January 6	& 2003 April 25 \\\hline
\multicolumn{3}{l}{{\bf Model 1: {\tt WA * PO}} $^a$} \\
\nh $^b$	& $0.56 \pm 0.08$	& $1.01^{+0.10}_{-0.12}$\\
$\Gamma$	& $2.14^{+0.05}_{-0.04}$	& $2.34^{+0.07}_{-0.06}$\\
$A_{\rm P}$ $^c$	& $(4.15 \pm 0.16) \times 10^{-4}$	& $(6.76^{+0.36}_{-0.34}) \times 10^{-4}$ \\
$\chi^2/$dof	& $573.6/540$	& $444.6/394$ \\
\multicolumn{3}{l}{{\bf Model 2: {\tt WA * BMC}}} \\
\nh		& 0 $^f$	& $0.41^{+0.18}_{-0.17}$ \\
$kT_{col}$ $^d$& $0.16 \pm 0.01$	& $0.17 \pm 0.02$\\
$\alpha$ $^e$	& $0.96 \pm 0.06$	& $1.09^{+0.12}_{-0.11}$ \\
log ($f$) $^g$	& $0.48^{+0.14}_{-0.13}$& $0.16^{+0.15}_{-0.14}$ \\
$A_{\rm BMC}$ $^c$	& $(1.1 \pm 0.1) \times 10^{-5}$	& $(1.8^{+0.4}_{-0.2}) \times 10^{-5}$ \\
$\chi^2/$dof	& $530.6/539$	& $380.6/392$ \\
\multicolumn{3}{l}{{\bf Model 3: {\tt WA * (DISKBB + PO)}}} \\
\nh	& $0.78^{+0.27}_{-0.17}$ 	& $1.22^{+0.34}_{-0.29}$ \\
$kT_{in}$ $^h$	& $0.21 \pm 0.03$ 	& $0.21^{+0.04}_{-0.03}$ \\
$A_{\rm D}$ $^c$ 	& $17^{+36}_{-10}$ 	& $45^{+88}_{-28}$ \\
$\Gamma$	& $1.97 \pm 0.07$	& $2.09^{+0.12}_{-0.13}$ \\
$A_{\rm P}$	& $(3.4^{+0.4}_{-0.3}) \times 10^{-4}$	& $(4.7^{+0.7}_{-0.7}) \times 10^{-4}$ \\
$\chi^2/$dof	& $530.9/538$	& $387.3/392$ \\
\multicolumn{3}{l}{{\bf Model 4: {\tt WA * (PO + DISKBB)}}} \\
\nh	& $1.68^{+0.48}_{-0.28}$ 	& $2.00^{+0.45}_{-0.34}$ \\
$\Gamma$	& $3.34^{+0.43}_{-0.26}$ 	& $3.34^{+0.41}_{-0.33}$ \\
$A_{\rm P}$	& $(4.7^{+0.6}_{-0.4}) \times 10^{-4}$ 	& $(8.3^{+1.0}_{-0.7}) \times 10^{-4}$ \\
$kT_{in}$	& $2.15^{+0.23}_{-0.25}$	& $2.78^{+1.65}_{-0.66}$ \\
$A_{\rm D}$	& $(2.2^{+1.6}_{-0.8}) \times 10^{-3}$	& $(8.4^{+17.4}_{-7.1}) \times 10^{-4}$ \\
$\chi^2/$dof	& $526.5/538$ 	& $409.6/392$ \\
  \\
$f_{\rm X}$ $^i$	& $1.6 \times 10^{-12}$	& $2.0 \times 10^{-12}$ \\
$L_{\rm X}$ $^j$	& $4.4 \times 10^{39}$	& $5.5 \times 10^{39}$ \\
\hline
\end{tabular} 
Notes: $^a$ Models are abbreviated to {\small XSPEC} syntax.  The
components are: {\tt WA} - cold absorption; {\tt PO} - power-law
continuum; {\tt BMC} - bulk motion Comptonization spectrum (see text);
{\tt DISKBB} - MCDBB.  $^b$ Absorption column external to our Galaxy
in units of $10^{21} \atpcm$.  A Galactic absorption column of $1.5
\times 10^{20} \atpcm$ is assumed (see text).  $^c$ {\small XSPEC}
model normalisations.  $^d$ Disc colour temperature in units of keV.
$^e$ Power-law spectral index.  $^f$ Value fixed at that shown.  $^g$
Illumination parameter (see e.g. Borozdin et al. 1999).  $^h$ Inner
disc temperature in keV.  $^i$ Flux from Model 4 (which agrees with
Models 2 and 3 to $\sim 3\%$), in units of $\ergcms$ (0.3 -- 10 keV).
$^j$ Derived X-ray luminosity, in $\ergsec$, for a distance of 4.8
Mpc.
\label{specfits}
\end{table}

The spectra from all three EPIC detectors were fit concurrently in the
0.3 -- 10 keV range.  We allowed for differences between the detectors
by including a constant multiplicative factor in each model, that was
fixed at unity for the pn data and allowed to vary for the two MOS
detectors.  Typical values obtained were 0.88/0.86 and 0.93/0.88 in
the first/second observations for MOS1 and MOS2 respectively, implying
a disagreement of $< 15\%$ in the absolute flux measurements from the
instruments (we quote an average of the three measurements where we
give fluxes).  We applied a fixed column of $1.5 \times 10^{20}
\atpcm$ to each spectral model to represent the foreground column
through our galaxy (interpolated from Stark et al. 1992), plus a
second absorption component left free to fit the data, which
represents additional absorption within NGC 5204 and/or intrinsic to
the ULX.

\begin{figure}
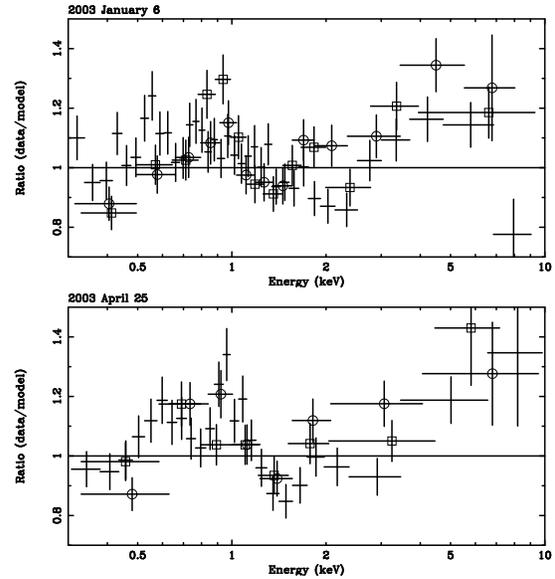

\centering
\includegraphics[width=3.8cm,angle=270]{fig2a.ps}
\includegraphics[width=3.8cm,angle=270]{fig2b.ps}
\caption{Residuals to the power-law continuum fits in both epochs.
The spectral data are rebinned to aid clarity, and data from all three
EPIC detectors are shown with MOS1/MOS2 data points identified by open
circles/squares.}
\label{residuals}
\end{figure}

The spectra appeared relatively smooth and featureless, hence we began
by fitting a number of simple continuum models to the data.  First we
tried a power-law continuum model; this gave reasonable fits, with
$\chi^2_{\nu} \sim 1.06$ and $\sim 1.12$ in the first and second
observations respectively.  The details of these fits are given in
Table~\ref{specfits} (Model 1).  We note that the overall spectrum of
NGC 5204 X-1 appears to have hardened substantially in the two years
between the \chan and \xmm observations, from $\Gamma \sim 2.8 - 3$ to
$\sim 2.15 - 2.35$ (cf. Roberts et al. 2004).  However,
Figure~\ref{residuals} demonstrates that the same substantial
systematic residuals to the power-law fit were present in both \xmm
observations, implying that more complex models are required to fit
the data.  In particular the spectral shape is inconsistent with any
single component model with a convex continuum form, such as a
classical blackbody, a thermal bremsstrahlung model, or the
``multi-colour disc blackbody'' spectrum used to represent the
optically-thick thermal emission from an accretion disc (hereafter
MCDBB; see e.g. Mitsuda et al. 1984).  As expected the spectral fits
of these ``single component'' models were rejected at a high
statistical significance in both sets of data ($\chi^2_{\nu} \gg 2$
for the blackbody and MCDBB, and $\chi^2_{\nu}
\ga 1.4$ for the thermal bremsstrahlung model).

We next considered models previously employed to explain the spectra
of ULXs.  A broken power-law continuum fit (e.g. Kaaret et al. 2003)
could represent X-ray emission produced in the Compton-scattering of
photons from the accretion disc and/or the companion star by a
relativistic jet (c.f. Georganopoulos, Aharonian \& Kirk 2002).
However, we ruled this model out as it predicts a spectral downturn at
low energies (i.e. $\Gamma_{low E} < \Gamma_{high E}$), whereas both
our spectra show an upturn at $\sim 1$ keV
(c.f. Figure~\ref{residuals}).  A second possible model was a
Comptonized disc model.  Kubota, Done \& Makishima (2003) suggest that
ULXs with a power-law-like X-ray spectrum are actually in a state
analogous to the very high state of Galactic black hole binaries,
where their emission is dominated by a strongly-Comptonized accretion
disc.  We first approximated this spectrum using the ``{\tt COMPTT}''
model of Titarchuk (1994) in {\small XSPEC}, and found better fits
than the power-law continuum model in both cases ($\chi^2_{\nu} \sim
1$).  However, the absence of data above 10 keV leaves the parameters
of this model poorly constrained - reasonable values (including a cool
130 -- 140 eV seed photon temperature) were only obtained after fixing
the electron temperature of the scattering medium at 50 keV.  As a
second approximation of a Comptonized disc spectrum we used the bulk
motion Comptonization model (``{\tt BMC}'', see Schrader \& Titarchuk
1999 and references therein), which has the twin advantages that it
does not require hard ($> 10$ keV) X-ray data to constrain its
fitting, and it provides an estimate of the inner accretion disc
temperature.  This model again provides a very good fit to each
dataset (Model 2 in Table~\ref{specfits}).  Most importantly, it
suggests that a ``cool'' accretion disc, with a temperature of $\sim
160 - 170$ eV, underlies the spectrum of NGC 5204 X-1.  This is
consistent with the presence of an accreting IMBH.

To confirm this result, and place it in the context of other recent
ULX observations, we then fit the data with a combined MCDBB +
power-law model.  This empirically describes the spectra of black hole
X-ray binaries (BHBs; cf. McClintock \& Remillard 2004; hereafter
MR04), and has been used to demonstrate evidence for cool accretion
discs in ULXs (e.g. Miller et al. 2003)\footnote{This description is
in essence a simplified version of the BMC model, in that it employs
the same spectral components, but it does not treat their relative
fluxes in a physically self-consistent manner, or truncate the
power-law continuum at low energies.}.  We find, as expected given the
Comptonization fits, that the data are very well fit by this
``standard'' model, displaying a cool ($kT \sim 0.2$ keV) MCDBB
component plus a hard power-law continuum.  The best-fit parameters
are listed in Table~\ref{specfits} as Model 3.  In each observation
the detection of the ``cool'' disc component is apparently very
robust, with F-test significances of $> 6\sigma$ and $> 7\sigma$
respectively when compared to the simple power-law continuum fit.

\begin{figure}
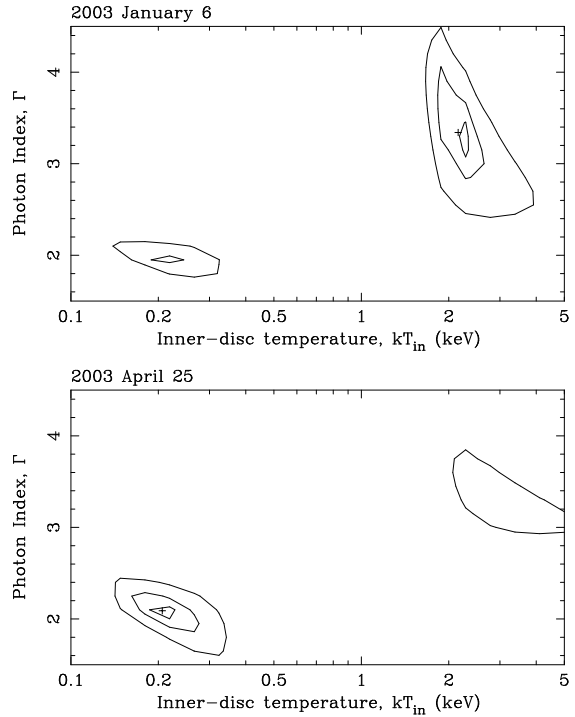

\centering
\includegraphics[width=4.6cm,angle=270]{fig3a.ps}\vspace*{0.2cm}
\includegraphics[width=4.6cm,angle=270]{fig3b.ps}
\caption{The minima in $\chi^2$-space, corresponding to Models 3 \&
4, shown as a function of $kT_{in}$ and $\Gamma$.  The contours are
displayed at $\Delta \chi^2 = 27, 9$ and 3 above the best fit for the
observation (with the largest deviation mapped by the outermost
contour).}
\label{ktvsgamma}
\end{figure}

Stobbart et al. (2004) have recently demonstrated that a variant of
the standard model provides the best empirical description of the
spectrum and spectral changes exhibited by a marginal (L$_{\rm X}
\approx 10^{39} \ergsec$) ULX in the very nearby galaxy NGC 55.  In
their analysis, the ULX spectrum changes over the course of two
contiguous 30-ks observations, initially having a soft power-law
continuum form ($\Gamma \sim 4$).  As the observation progressed a
second, curved continuum spectral component was required to model an
increase in the flux predominantly above 1 keV, that was well-fitted
by a $kT_{in} \sim 0.9$ keV MCDBB spectrum.  Whilst utilising the same
pair of components as in the standard description, this particular
model is distinguished by the power-law continuum dominating {\it at
low energies\/}, the reverse of the usual prescription. [A caveat, of
course, is that this strictly applies only to observations carried out
in the limited 0.3 -- 10 keV bandpass of {\it XMM-Newton\/}.]  This
``non-standard'' spectral model was also independently established to
provide a good fit for the nearest persistent extra-galactic ULX, M33
X-8, with $\Gamma \sim 2.5$ and $kT_{in} \sim 1.16$ keV, by Foschini
et al. (2004).  The NGC 55 and M33 X-8 ULXs have L$_{\rm X} \sim 1 - 2
\times 10^{39} \ergsec$ and, therefore, are both at the low-luminosity
end of the ULX regime. The question of whether the non-standard model
provides at least an empirical representation of the spectra of more
luminous ULXs therefore merits some consideration.  On this basis we
have attempted to fit the non-standard model to the current data.  In
fact an acceptable fit to the data from both epochs was obtained by
this approach (Model 4 in Table~\ref{specfits}).  We explicitly
demonstrate that the data allows two $\chi^2$ minima - corresponding
to Models 3 \& 4 - by plotting the minima in $kT_{in} - \Gamma$
confidence space in Figure~\ref{ktvsgamma}.  

\begin{figure}
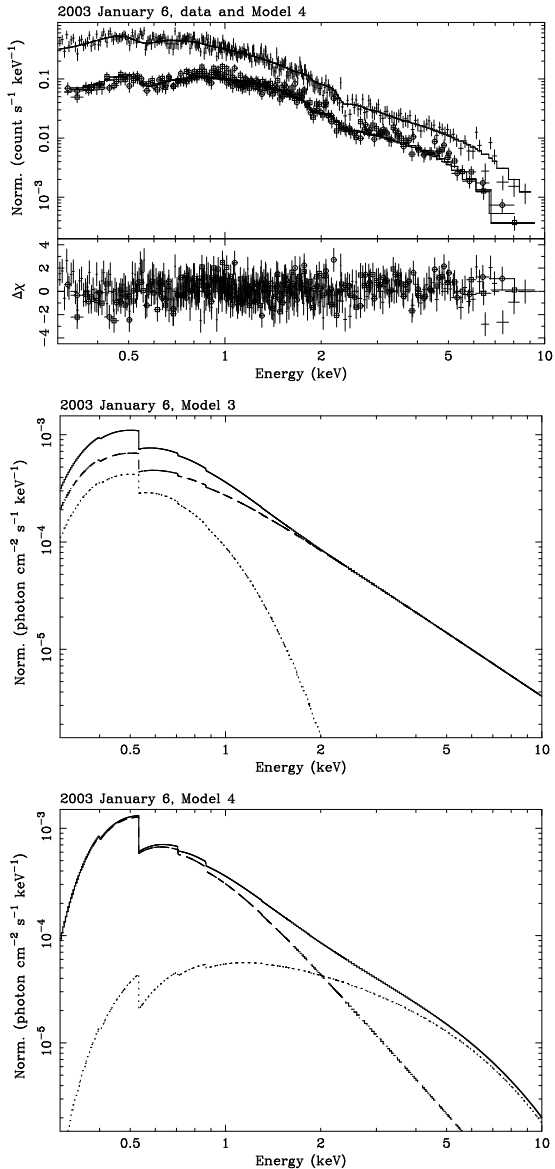

\centering
\includegraphics[width=5cm,angle=270]{fig4a.ps}\vspace*{0.2cm}
\includegraphics[width=5cm,angle=270]{fig4b.ps}\vspace*{0.2cm}
\includegraphics[width=5cm,angle=270]{fig4c.ps}
\caption{Best fit spectra to the first observation using the variants of
the MCDBB + power-law continuum model.  ({\it top panel\/}) The data (with
MOS1/MOS2 points distinguished by open circles/squares as before) and
best fit Model 4 spectrum (solid line) are shown in the upper part of
the panel, and the fit residuals in $\Delta \chi$ space in the lower
part.  ({\it centre \& bottom panels\/}) A comparison of the variants of the
MCDBB + power-law continuum models.  The overall fit is shown as a
solid line, with the contributions from the MCDBB and power-law
continuum distinguished as dotted and dashed lines respectively.}
\label{specplots}
\end{figure}

In Figure~\ref{specplots} we show the data and best-fit model (Model
4) for the first observation (the fit quality in the second
observation is consistent with that shown), and include a direct
comparison of the underlying forms of Models 3 \& 4.  The composite
spectrum is almost identical in the two cases, with the largest
divergence evident at high energies ($> 3$ keV), where Model 4 shows a
spectral downturn due to the distinct curvature in the dominant MCDBB
component.  We have further investigated whether the data best
supports Model 3 or Model 4 by explictly searching for evidence of
this high-energy break in the spectral data.  We restricted our
analysis to the data points above 2 keV, and modelled the spectra
using both single and broken power-law continua.  We find that a
broken power-law is not required by the second (April 25) dataset,
which is very well-modelled above 2 keV by a $\Gamma \sim 2.1$
power-law continuum ($\chi^2$/dof = 84.9/83), consistent with Model 3.
However, whilst the first (January 6) dataset is also well-fitted by a
$\Gamma \sim 2.45$ power-law continuum above 2 keV ($\chi^2$/dof =
137.4/136), we find that the fit is improved at the $\sim 3\sigma$
confidence level (according to the F-statistic, i.e. $\Delta \chi^2 =
9.7$ for two additional degrees of freedom) by a broken power-law fit
with $\Gamma_1 = 1.8^{+0.4}_{-0.3}$, E$_{break} = 4.9^{+0.5}_{-0.4}$
keV, and $\Gamma_2 = 3.1^{+0.6}_{-0.5}$.  Hence the first observation
shows marginal evidence for the presence of a spectral downturn at 5
keV, which (in combination with the slightly better fit to the full
spectrum) favours Model 4 as the best description of its spectral
shape.

\section{Discussion}

\subsection{IMBH models}

Both the Comptonization models and the accepted empirical model for
BHB spectra point towards the presence of a cool accretion disc
signature in the X-ray spectrum of NGC 5204 X-1.  As discussed by
previous authors (e.g. Miller et al. 2003), this suggests that an IMBH
is present in this system.  Indeed, using equations (5) and (8) of
Makishima et al. (2000), the best-fit MCDBB parameters translate into
masses of $256 ({\rm cos}(i))^{-{{1}\over{2}}}$ and $412 ({\rm
cos}(i))^{-{{1}\over{2}}}$ M$_{\odot}$ (where $i$ is the inclination
of the accretion disc with respect to the observer) for the first and
second observations respectively. Clearly this places the accreting
object well into the IMBH regime\footnote{In this calculation we
assume $\alpha = 1$ in equation (8) of Makishima et al. (2000),
i.e. the Schwartzschild metric.  A Kerr metric would imply $\alpha <
1$ hence a larger black hole mass.}.

The optical counterpart of NGC 5204 X-1 (HST-3 of Goad et al. 2002)
lies in close proximity to a probable young stellar cluster, HST-1.
Liu et al. (2004) resolve HST-1 into a point-like red component (a
possible young cluster of a few thousand stars) and a more diffuse
blue component, probably an OB-association.  This proximity is
potentially important given recent results suggesting that IMBHs can
be formed by runaway stellar collisions in young, dense stellar
clusters (Portegies-Zwart et al. 2004 and references therein).  The
projected separation of HST-3 from HST-1 is only $\sim 26$ pc; if the
ULX originated in the cluster, this distance must have been traversed
in less than the $\sim 10^7$ year lifetime of the B0 Ib secondary.
This is possible with a transverse velocity as small as $\sim 2
\kmsec$ for the ULX relative to the stellar cluster.  If we consider
that stellar-mass black holes in our Galaxy are seen with spatial
velocities, presumably imparted by a kick at the time of the black
hole formation, of the order $\sim 100 \kmsec$, then the minimum
velocity requirement appears entirely plausible for an ULX harbouring
a $\sim 500$ M$_{\odot}$ IMBH, formed in, and subsequently kicked-out
of the cluster (cf. Zezas et al. 2002 for similar calculations for
ULXs in the Antennae).

However, there is an anomaly in the parameterization of spectral Model
3, from which the IMBH result is derived.  The detection of an IMBH is
based on a reasonable extrapolation of stellar-mass black hole
properties, as represented by the empirical MCDBB + power-law spectral
model, to the IMBH regime, with the ``cool'' accretion disc signature
the key evidence for an IMBH.  Because of the extreme luminosities of
ULXs it is generally assumed that the accretion disc is in a ``high''
state, where it extends to the innermost circular orbit around the
black hole.  This must correspond to the classic ``high/soft'' state
in a stellar-mass BHB (also known as the ``thermal dominated'' state,
cf. MR04).  However, in BHBs the MCDBB in this state is typically
accompanied by a power-law component with a measured photon index $>
2.4$.  This is also true for the ``very high'' (steep power-law)
state.  In fact, such steep slopes are theoretically expected from
Comptonized accretion disc spectra of sources close to the Eddington
limit (Chakrabarti \& Titarchuk 1995).  Assuming that the \xmm
passband allows a reasonable measurement of the power-law slope, then
such steep slopes are not present in our data when parameterized by
Model 3.  This anomaly is actually seen in several ULXs with inferred
soft accretion discs (e.g. Miller et al. 2003; Miller, Fabian \&
Miller 2004), though not all of them (cf. Soria et al. 2004; Dewangen
et al. 2004).  The anomalous power-law slopes derived for ULXs with
the IMBH model are generally $\la 2$, which is actually more
consistent with the ``low/hard'' state in BHBs.

So could some ULXs instead be low/hard state sources?  This state
generally shows much cooler accretion disc temperatures than the
high/soft state in BHBs (possibly due to the truncation of the inner
edge of the accretion disc at large radii from the black hole; MR04),
with observed values of $kT_{in}$ similar to the temperatures derived
in the IMBH model.  Hence a low temperature disc is not necessarily an
unambiguous indicator of the presence of an IMBH. [We note that an
IMBH in the low/hard state would, by extended analogy, have an even
cooler disc than that observed ($kT_{in} \sim 10$s of eV).]  However,
the luminosities associated with the low/hard state in BHBs are
generally very low by comparison to ULXs (\lx ~$\sim 10^{36} \ergsec$).
This characteristic lack of X-ray luminosity appears to rule out
stellar-mass black holes in the low/hard state as candidates to
explain the power-law slope of Model 3.  It seems therefore that the
classic BHB states may not always be exact analogies of ULX behaviour,
which could be problematic for the IMBH interpretation.

\subsection{``Non-standard'' models}

An alternative description of the X-ray spectrum of NGC 5204 X-1 is
provided by Model 4, which as previously stated has been shown to
provide the best-fit model to high-quality \xmm spectra in two ULXs
(Stobbart et al. 2004; Foschini et al. 2004).  In this description,
both the inner-disc temperature and power-law slope are well within
the range of known behaviour for the high/soft and very high states in
Galactic stellar-mass BHBs ($kT_{in} > 0.7$ keV; $\Gamma >
2.5$)\footnote{Indeed, the derived values in observation 1 are
extraordinarily close to the $kT_{in} \sim 2.2$ keV, $\Gamma \sim 3.5$
fit quoted for GRS 1915+105 in its high/soft state by MR04.}.  Whilst
the inferred disc temperatures for NGC 5204 X-1 in this model do
appear high for a stellar-mass system, similar temperatures have been
observed from BHBs in the high/soft and very high states (MR04) and,
interestingly, a high disc temperature is consistent with the ``slim
disc'' regime for near- and slightly super-Eddington accretion
(Watarai, Mizuno \& Minishige 2001; Ebisawa et al. 2003).  A
stellar-mass system is also not ruled out from the proximity of the
ULX to a possible young stellar cluster.  Indeed, a very plausible
scenario is that the ULX is simply a high-mass X-ray binary recently
formed within, and kicked-out of, the OB-association.

However, there are physical problems with the non-standard model.  The
MCDBB normalisation translates into inner disc radii ($R_{in}$) of
$\sim 26.5$ and $\sim 16$ km.  These are very problematic as they are
somewhat smaller than the 30 km Schwarzschild radius of a
10-M$_{\odot}$ black hole.  However, comparable discrepancies are also
seen in some Galactic systems.  For example, Ueda et al. (2002) use
\asca data to measure $R_{in} \sim 30$ km for two observations of GRS
1915+105, a black hole that has a dynamically-determined mass of $14
\pm 4$ M$_{\odot}$ (Greiner, Cuby \& McCaughrean 2001).  Again,
apparently small inner-disc radii are also a prediction of the slim
disc model (Watarai, Mizuno \& Minishige 2001).  We also note the
apparent small inner-disc radius in NGC 5204 X-1 could in part be
explained by an inclination $i > 0$ which will serve to increase
$R_{in}$.  The masses derived from the MCDBB model normalization are
within the stellar-mass black hole regime, at $3 ({\rm
cos}(i))^{-{{1}\over{2}}}$ and $1.8 ({\rm cos}(i))^{-{{1}\over{2}}}$
M$_{\odot}$ for the two observations.  This suggests that NGC 5204 X-1
could contain a stellar-mass black hole.  For a $\sim 10$ M$_{\odot}$
black hole, this would require a radiation rate at $\sim 4 - 5$ times
the Eddington luminosity.  We note this is well within the predictions
of Begelman (2002) and King et al. (2001) for truly
super-Eddington/apparently super-Eddington accretion discs, and such
luminosities have been observed (albeit briefly) from Galactic
stellar-mass systems (MR04).

The most severe problem with the non-standard spectral description is
the domination of the power-law continuum at low energies.  This
spectral form has not (to the best of our knowledge) previously been
observed for any Galactic BHB.  Furthermore, it is physically
inconsistent with the standard accretion disc plus Comptonized-corona
model for BHBs, as one would not expect to see the power-law continuum
extend below the peak emissivity of the MCDBB component, due to a lack
of seed photons available to be Compton-scattered.  It is very
unlikely that the corona is simply seeing an alternate source of seed
photons; for instance it would require a radically different disc
emission profile for these photons to originate in the cool outer
regions of the accretion disc.  It is also very unlikely that the
secondary star could seed this component; assuming equal primary and
secondary masses of $\sim 10$ M$_{\odot}$, with the B0 Ib secondary
overflowing its Roche lobe (cf. Liu et al. 2004), and a corona
extending out to 100 Schwarzschild radii, then the coronal region
intercepts less than one in a million photons emitted by the star.  As
the power-law component emits $\sim 10^{49}$ photon s$^{-1}$ above 0.3
keV (after correcting for absorption) this requires implausibly high
stellar fluxes.

So can the dominant soft power-law continuum instead be explained by
an alternative physical process?  Foschini et al. (2004) discuss two
possibilities for M33 X-8; mildly-relativistic jets, and an outflowing
wind.  A mildly-relativistic jet is certainly a possibility in M33 X-8
due to the detection of a radio counterpart.  However, Galactic black
holes tend to display spectrally-hard power-law continua ($\Gamma \sim
1.5 - 2$) during episodes of radio jet emission, predominantly when in
the low/hard spectral state (though jets can also be seen from very
high state sources, e.g. GRS 1915+105; Reig, Belloni \& van der Klis
2003).  Such a hard power-law continuum is inconsistent with the X-ray
spectral forms seen in these ULXs.  The alternative, an outflowing
wind, may be likely if ULXs are powered by accretion from a high-mass
star onto a stellar-mass black hole.  For instance, Podsiadlowski,
Rappaport \& Han (2003) demonstrate that the rate of matter transfer
from the star could potentially be far higher than that required to
reach the Eddington limit in these systems.  Presumably this excess
material must either be accreted onto the black hole in a
radiatively-inefficient manner, or more likely discharged from the
system in an outflow (see Begelman 2002, King 2002).  This scenario
may be particularly appropriate for NGC 5204 X-1, which possesses a B0
Ib star companion (Liu et al. 2004).  However, whilst such a wind
could emit a soft X-ray spectrum (e.g. King \& Pounds 2003), it should
have a thermal blackbody spectrum.  Hence this also appears to be
unable to provide an origin for the dominant soft power-law.

The lack of an obvious explanation for the origin of the soft
power-law continuum seems to provide insurmountable physical problems
for this model.  However, given that it provides a clear description
of the shape of the X-ray spectra of M33 X-8 and the NGC 55 ULX (and
is marginally preferred in the first observation of NGC 5204 X-1),
Model 4 must demonstrate at least subtle differences to, or perhaps
greater complexity than, the physical processes underlying in the
standard accretion disc plus optically-thin corona model.  We
speculate that one physical picture that may explain this distinct
spectral shape is presented by Zhang et al. (2000).  They describe
accretion discs around stellar-mass black holes by a three-layer model
roughly analogous to the Solar corona, composed of a cold (0.2 - 0.5
keV) inner-disc layer, a warm and optically-thick atmosphere to the
disc ($\tau \sim 10$, $kT \sim 1 - 1.5$ keV) and a much hotter
optically-thin corona.  We note that such a model will inevitably
produce a Comptonized component down to the energies suggested by the
soft power-law continuum in Model 4.  Crucially the warm,
optically-thick accretion disc atmosphere will also produce a spectral
turnover at energies of a few keV, similar to that we have modelled by
a MCDBB.  It is therefore possible that at least some ULXs could have
X-ray spectra dominated by this optically-thick warm accretion disc
atmosphere.  This will be investigated in detail in future work.

\subsection{Concluding comments}

These observations imply a degree of spectral ambiguity as models
supporting the presence of an IMBH (Models 2 and 3 of
Table~\ref{specfits}) and the non-standard alternative (Model 4)
provide equally acceptable fits to the available data.  In our view
the balance of the argument still appears to favour the presence of an
IMBH in NGC 5204 X-1, although its hard power-law slope remains as
somewhat of an anomaly.  However the contrasting interpretation of the
data discussed in this paper clearly merits further investigation.

\vspace{0.2cm}

{\noindent \bf ACKNOWLEDGMENTS}\\ We thank the second referee for
their fair and balanced review of this paper.  TPR and LPJ gratefully
acknowledge support from PPARC.  This work is based on observations
obtained with {\it XMM-Newton\/}, an ESA science mission with
instruments and contributions directly funded by ESA member states and
the USA (NASA).

\end{document}